\documentclass[preprint,preprintnumbers,amsmath,amssymb]{revtex4}
\usepackage{graphicx}

\makeatletter

\newenvironment{figurehere}
  {\def\@captype{figure}}
  {}
\makeatother

\begin{document}
\title{Electronic Properties of
Carbon Nanotubes Calculated from Density Functional Theory
and the Empirical $\pi$-Bond Model}
\author{Deep Shah}
\author{Nicolas A. Bruque}
\email{nbruque@ee.ucr.edu}
\author{Khairul Alam}
\author{Roger K. Lake}
\email{rlake@ee.ucr.edu}
\author{Rajeev R. Pandey}

\affiliation{Department of Electrical Engineering, University of
California, Riverside, CA 92521-0204 \\
J. Computational Electronics (accepted, 20 February 2007)}

\begin{abstract}
The validity of the DFT models implemented by FIREBALL
for CNT electronic device modeling is assessed.
The effective masses, band gaps, and
transmission coefficients of semi-conducting, zigzag, $(n,0)$ carbon
nanotubes (CNTs) resulting from the \emph{ab-initio} tight-binding
density functional theory (DFT) code F{\footnotesize IREBALL} and
the empirical, nearest-neighbor $\pi$-bond model are compared
for all semiconducting $n$ values $5 \leq n \leq 35$.
The DFT values for the effective masses differ
from the $\pi$-bond values by $\pm 9$\% over
the range of $n$ values, $17 \leq n \leq 29$, most important for electronic device applications.
Over the range $13 \leq n \leq 35$,
the DFT bandgaps are less than the empirical bandgaps by 20-180 meV depending
on the functional and the $n$ value.
The $\pi$-bond model gives results that differ significantly from
the DFT results when the CNT
diameter goes below 1 nm due to the
large curvature of the CNT.
The $\pi$-bond model quickly becomes inaccurate away from the bandedges
for a $(10,0)$ CNT, and it is completely inaccurate for $n \leq 8$.
\\
\\
{\em Keywords}: F{\footnotesize IREBALL}, CNT, DFT, NEGF,
Bandstructure.
\end{abstract}
\maketitle

\section{INTRODUCTION}

Carbon nanotube (CNT) systems are of high research interest for use
in sensing and nanoelectronics.
Electronic device and
circuit architecture concepts for bio-assembled CNTs
have been described and demonstrated
\cite{Braun_Science03,Braun_Adv_Phys04,Dekker_Nature_DNA_PNA02,Dekker_PNA04,Dwyer_Design_tools04}.
One system that we have found particularly interesting is a
molecule joining two semiconducting CNTs.
Such a system can display the electrical response of
a resonant tunnel diode (RTD)
\cite{Bruque_Pandey_MolSim,Pandey_PSSA_MolRTD06,JNO_v1_06,Xu_CNT_DNA_Small06}, and we
refer to it as a CNT-Mol-RTD.

The current-voltage response of a CNT-Mol-RTD depends
on the alignment of the CNT bandedges with
molecular states of the organic group
\cite{Bruque_Pandey_MolSim,Pandey_PSSA_MolRTD06,JNO_v1_06}.
Therefore, quantitative simulations of such a structure
require models which are suitable for calculating the
electronic states of both semiconductors,
molecules, and the chemical bonds between the molecule
and semiconductor. The models should be
general enough to treat local distortions in the
semiconductor lattice and distortion of the molecule.
A CNT-Mol-RTD structure
tends to be large compared to the benzene dithiol type of
molecules that have been so heavily studied for molecular electronics
\cite{DAMLE:PRB:2001,DAMLE:CP:2002,RAKSHIT:NL:2004,XUE:CP:2002,
XUE:JCP:2001,YALIRAKI:JCP:1998,SEMINARIO:IEEE:2003}.
Therefore, the model and its implementation must have the
ability to handle large systems.

One model widely used for computational molecular electronics is
density functional theory (DFT).
DFT is a general, flexible theory with a proven record
of calculating the electronic states of both semiconductors
and molecules. The well known weakness of DFT implemented
with widely used density functionals is its
underestimation of the bandgap of semiconductors \cite{RichardMartin}.
For electronic device modeling, this is not a trivial issue.
The most famous example is that of Ge which is predicted
within the local density approximation (LDA)
or generalized gradient approximation (GGA) to be a metal.
Another example is Zener tunneling which is exponentially dependent on the
bandgap. It is this process that
limits the maximum on-off current ratio
in CNT field effect transistors.
Linear errors in the bandgap result in
exponentially large errors in the simulated minimum
off-current \cite{khairul_APL1}.
The final example is the electron current of a CNT-Mol-RTD.
It depends on the alignment of a molecular state with
the conduction band edge of the CNT.
If the conduction band edge is too low, the predicted current - voltage
response will be inaccurate.
While hybrid functionals that include exact exchange greatly improve the situation,
they also significantly increase the computational burden \cite{bandgaps_B3LYP}.

Another option for the simulation of large structures is to use
empirical models \cite{Zunger_Comp_Phys00,klimeck_CMES02}.
Such models can reproduce the bandgaps and energy levels of
semiconductors with high accuracy.
However, simulations of heterogeneous CNT, metal, organic, biological systems
are difficult
with empirical models due to
issues of transferability of parameters.
Conversely,
electronic device simulations of semiconductors are difficult with
DFT models due to the underestimation of the bandgap.
One is forced to make a choice, and we have chosen
the DFT model as implemented by the code F{\footnotesize IREBALL}
\cite{Sankey89,Further_developments} since it has
demonstrated the ability to model large biological molecules \cite{States_DNA03}.

In this paper, we compare predictions of the properties relevant to
electronic device modeling of semiconducting (n,0) CNTs calculated
from both DFT theory and an empirical, tight-binding, $\pi$-bond
model. Two important band-edge properties are the bandgap and the
effective masses. For larger diameter CNTs, the empirical,
tight-binding model is expected to predict the bandgap and band-edge
effective masses with good accuracy since the empirical parameters
have been chosen by fitting to those quantities. Therefore, for the
larger diameter CNTs, we use the values of the bandgap and effective
masses obtained from the $\pi$-bond model to assess the values
obtained from the DFT model. For smaller diameter CNTs where the
curvature becomes significant, the $\pi$-bond model breaks down.

Bandedge quantities alone are not sufficient for electronic device
modeling.
Since electronic devices are operated at biases on the order of a volt,
accurate modeling of
the higher energy states away from the bandedges is
necessary for device modeling.
The quantity that best characterizes the higher energy spectra
for device simulations is the transmission coefficient.
Therefore, we compare transmission coefficients calculated from
the DFT and the empirical, tight-binding models.

Below, we calculate and compare the bandgaps, effective masses,
and transmission spectra of
semiconducting zigzag CNTs, ranging from $(5,0)$
to $(35,0)$ corresponding to diameters ranging from $0.39$ nm to
$2.8$ nm. The bandgaps and effective masses are calculated, plotted,
and compared
for every non-metallic $(n,0)$ CNT with $5 \leq n \leq 35$.
Selected transmission coefficients are plotted and compared
for $n\;=$ 10, 20, 31, and 35.

\section{METHOD OF CALCULATION}

The F{\footnotesize IREBALL} calculations are performed using the
local density approximation (LDA) (the Ceperley-Alder
\cite{Ceperly_Alder80} form as parameterized by Perdew and Zunger
\cite{Perdew_Zunger81}) and the BLYP exchange-correlation functional
\cite{Becke88,LYP88}. A self-consistent calculation is performed
using a generalization of the Harris-Foulkes \cite{Harris85,
Foulkes89} energy functional referred to as DOGS after the authors
of the original paper
\cite{Demkov_charge_transfer95,Approx_dens_fn98}. A separable
non-local pseudopotential \cite{Hamann89} and a minimal sp$^3$
F{\footnotesize IREBALL} basis are used. The localized pseudoatomic
orbitals are slightly excited due to hard wall boundary conditions
imposed at certain cutoffs, \emph{$r_c^{2s}$} and \emph{$r_c^{2p}$}
\cite{Approx_dens_fn98}. An excitation energy of approximately 2.0
eV is used to preserve the chemical bonding trends of carbon which
results in $r_c^{2s}$ = 4.0 {\AA}  and $r_c^{2p}$ = 4.5 {\AA}.
Further details are given in \cite{Further_developments}.

After the F{\footnotesize IREBALL} DFT calculation finishes, the
device Hamiltonian matrix elements,
the overlap matrix,
and the device-to-contact coupling matrices are extracted.
The spatial extent of the non-zero
matrix elements (the sparsity of the matrices) is determined by the
pseudopotential cutoff limits and the F{\footnotesize IREBALL}
orbital radii.  Each CNT system, pictured in
Fig. \ref{fig:CNT_SCHEMATIC}, consists of one CNT unit cell composed of 4
atomic layers periodically repeated in the axial direction.
Non-zero matrix elements of a given atomic layer
extend to the left and right 4 atomic layers, or one unit cell of
the zigzag CNT.
In terms of the 4-atomic layer unit cells, there is only
nearest neighbor unit-cell coupling.

Transmission is calculated using the non-equilibrium Green function
formalism (NEGF).
The CNT is partitioned into a `device' consisting of one unit cell
and a left and right `contact.'
The left and right `contacts' are taken into account exactly by self-energies
$\Sigma^{\ell}$ and $\Sigma^r$, respectively,
as illustrated in Fig. \ref{fig:CNT_SCHEMATIC}.
Details of the NEGF algorithm are described in
\cite{Bruque_Pandey_MolSim,Pandey_PSSA_MolRTD06,JNO_v1_06}.

For the empirical $\pi$-bond model, we use a nearest-neighbor model
with matrix element $V_{pp\pi} = -2.77$ eV and $\epsilon_p$ = 0.0 eV
\cite{MINTMIRE:JPCS:93}. The NEGF algorithm is the same as that used
with the F{\footnotesize IREBALL} matrix elements. The effective
mass for both models is calculated from the 1-D dispersion using
$1/m* = \frac{1}{\hbar^2}
\partial^2 E / \partial k^2$.
The energy band gap is determined from
an $E-k$ calculation by
reading the difference between the highest occupied band
energy and the lowest unoccupied band energy at
the $\Gamma$ point.

\section{RESULT AND DISCUSSION}

Figure \ref{fig:bandgap}(a) compares the calculated band gaps
for $(n,0)$ CNTs with $5 \leq n \leq 35$ leaving out the metallic CNTs with
$n$ values that are integer multiples of 3.
The $n$ values are shown
on the bottom horizontal axis and the corresponding CNT diameters
are shown on the top horizontal axis.
On the plot itself, the data points indicate
the $n$ values for which calculations were performed.
All other $n$ values in that range correspond to metallic CNTs.
At first glance, the bandgaps that result from the
FIREBALL calculations closely track the bandgaps determined by
the $\pi$-bond model for $n \geq 10$.
CNTs in this size range are the ones that are most
commonly synthesized, and they are the ones that are most important
for electronic devices \cite{Z.Chen_CNT_contacts}.
Below $n=10$,
the $\pi$-bond model breaks down due to the large curvature
of the CNT.
For the smallest CNT with $n=5$, the FIREBALL calculations
show zero bandgap.
Upon closer inspection of Fig. \ref{fig:bandgap}(a),
one notices a sawtooth shape to the plot of bandgap
versus $n$ for the DFT calculations.
We observe that the bandgap resulting from the DFT calculations
corresponds closely to the bandgap resulting from the
$\pi$-bond model for $(n,0)$ CNTs when $n = 3p + 1$ where $p$ is an integer
$\geq 3$.

This is shown clearly in Fig. \ref{fig:bandgap}(b).
in which we plot the differences between
the bandgaps calculated from the LDA and BLYP functionals
and those from the $\pi$-bond model for $n \geq 10$.
For $n = 3p+1 \geq 13$, the LDA model underestimates the bandgap by 24 - 28 meV.
For these $n$ values, the BLYP model underestimates the bandgap by 46 - 84 meV.
For $n = 3p - 1$ with $p$ an integer, the discrepancies between the bandgaps resulting from the DFT
models and the $\pi$-bond model are larger.
For values of $n = 3p-1 \geq 14$, the LDA model underestimates
the bandgap by 52 - 147 meV.
For these $n$ values, the BLYP model underestimates the bandgap by 66 - 176 meV.

Figs. \ref{fig:effM_comparison} shows calculations
and comparisons of the electron and hole effective masses.
The plots of the left column show calculations
and comparisons of the electron effective mass,
and the plots of the right column show calculations
and comparisons of the hole effective mass.
In each row consisting of two plots, the scale and range of values is
identical to facilitate easy comparisons between the
values for electrons and holes.
The calculated electron and hole effective masses
(normalized to the bare electron mass)
are plotted in Figs. \ref{fig:effM_comparison}(a) and (d), respectively,
versus $n$ for all semiconducting values in the range $7 \leq n \leq 35$.
One immediately notices that all models result in very similar values
of effective masses
for both the electrons and holes.
To discern the differences, we plot the difference of the mass
values calculated from the DFT models and the pi-bond model
in Figs. \ref{fig:effM_comparison}(b) and (e) for electrons and
holes, respectively.
In other words, Figs. \ref{fig:effM_comparison}(b) and (e) show
the normalized difference values
$(m^*_{\rm DFT} - m^*_{\pi-{\rm bond}})/m_0$.
Note that on the mass difference plots,
(b) and (e), the
$n$ values range from 10 - 35 whereas on the mass plots, (a) and (d),
the $n$ values range from 7 - 35. The range of $n$ is reduced
on the difference plots to keep the plotted range of differences
small.
The difference plots show that the effective mass
determined from the BLYP model tends to be
larger than the effective mass determined by the
LDA model.
Over the range of $n$ values $14 \leq n \leq 35$, the
mass determined by the BLYP model has a maximum difference
from the mass determined from the LDA model of 0.008 at
$n=14$ and a minimum difference of 0.002 at $n=34$.
Also, the sawtooth peaks in the difference curves are
out of phase with the sawtooth peaks in the mass curves
above. The maximum differences occur at the minimums of the
mass curves.
Overall, the differences in the mass values predicted by the
DFT models and the $\pi$-bond model are relatively small.
To quantify the differences, the percent differences,
$100 * (m_{\rm DFT} - m_{\pi-{\rm bond}}) / m_{\pi-{\rm bond}}$,
are plotted in (c) and (f).
Over a wide range of the most useful $n$ values for device
applications,
$17 \leq n \leq 29$, the values for the
effective masses from the DFT models
fall to within $+8$\% to $-9$\% of
the values from the $\pi$-bond model.

So far, we have considered bandedge properties of
bandgaps and effective masses. As we noted above,
the higher energy electronic spectra is also important
for electronic device modeling, and the best way
to characterize it is to calculate the
transmission coefficients.
Figure \ref{fig:TE_comparison} shows the transmission spectra for
$(10,0), (20,0), (31,0)$ and $(35,0)$ CNTs calculated using
DFT/BLYP, DFT/LDA, and the empirical, $\pi$-bond model. In all
cases, the energy axis of the transmission curves has been shifted
such that the center of the bandgap lies at 0 eV. The energy region
in Fig. \ref{fig:TE_comparison} for which the transmission is zero
is the band gap for the CNTs.

For the 2 largest CNTs, $n$ = 31 and 35,
the transmission resulting from the DFT and $\pi$-bond
models all have similar, symmetric forms.
There is some compression of the energy scale
for the transmission coefficients calculated
from the DFT models compared to the
transmission coefficient calculated from
the $\pi$-bond model. The energy separation between
higher modes is smaller
in the DFT models then in the $\pi$-bond model.

For the smallest CNT, $n$ = 10,
there is a noticeable, qualitative difference between the
transmission calculated from the DFT models
and the $\pi$-bond model.
The transmission resulting from the $\pi$-bond model
is always symmetric around the center of the bandgap.
For the DFT models, the transmission is
noticeably asymmetric.
Approximately 0.5 eV above the conduction band edge,
the DFT models predict
3 bands closely spaced and 2 bands doubly degenerate.
These 7 bands, multiplied by 2 for spin, give rise to
the large step of 14 in the transmission coefficient
0.5 eV above the conduction band edge
in the transmission coefficient of the
$(10,0)$ CNT.
This large increase in the transmission and
density of states 0.5 eV in the conduction
band is significant for device modeling.
A similar large step is also observed in a
transmission calculation based on the SIESTA code
for a $(7,0)$ CNT \cite{ADESSI:PRB:06}.
A similar large increase in the transmission
also occurs in the (20,0) CNT at 1.3 eV above the
conduction band edge.

The differences in the (10,0) valence band transmission
resulting from the DFT and $\pi$-bond models, while
not as dramatic as those in the conduction band,
are still significant from a device modeling
perspective.
The 0.5 eV gap between the valence band edge and
the next lower pair of bands found from the
$\pi$-bond model is reduced to approximately
0.3 eV in the DFT models.
These energies lie within the applied voltage window, $(V_{DD})$,
of any of the most optimistically scaled CNT
field effect transistors, and will, thus,
affect the physics of the carrier transport.
While our main focus is on assessing the validity of
the DFT models for device modeling, these results also
provide an assessment of the $\pi$-bond model and show that
the $\pi$-bond model should be used with care and scepticism
for (10,0) CNTs.

\section{SUMMARY}

The goal of this work is to assess the validity of
the DFT models implemented by FIREBALL for CNT electronic device modeling.
Our approach is to compare the electronic properties resulting
from the DFT models with those resulting from the $\pi$-bond
model since the parameters of the $\pi$-bond model
have been empirically chosen to give a good fit to
the bandgap and effective mass for CNTs with diameters
that are `not too small.'
We have compared the bandgaps, effective masses, and transmission
coefficients of $(n,0)$ CNTs calculated from the empirical
$\pi$-bond model, DFT/LDA, and DFT/BLYP models.

For values of $n$ in the range $17 \leq n \leq 29$,
the calculated effective masses from the DFT models are within
$\pm 9$\% of those calculated from the $\pi$-bond model.
For $n \geq 10$, the difference between the bandgap calculated from
the $\pi$-bond model and the bandgaps calculated from
the DFT models oscillates as a function of $n$.
The differences are smallest for $n = 3p+1$ where $p$
is an integer, and the differences are largest
for $n = 3p - 1$.
For $n = 3p+1 \geq 13$, the LDA model underestimates the bandgap
by 24 - 58 meV and the BLYP model underestimates
the bandgap by 46 - 84 meV.
For $n = 3p - 1 \geq 14$, the LDA model underestimates
the bandgap by 52 - 147 meV and the BLYP model
underestimates the bandgap by 66 - 176 meV.
Overall, in the important range of $n$ values most
relevant for CNT devices, $17 \leq n \leq 29$,
the bandgaps, effective masses, and transmission coefficients calculated from the DFT models implemented
by FIREBALL are sufficiently accurate for electronic
device simulations.

These simulations also quantify what is meant by
`not too small' when applying the $\pi$-bond model.
For $n = 10$, the bandedge properties resulting from the
$\pi$-bond and DFT models agree to within 10\%,
however, the $\pi$-bond model quickly becomes inaccurate
away from the bandedges.
The transmission from the higher energy modes
resulting from the
$\pi$-bond model has differences with those resulting
from the DFT models which are significant for device
modeling.
For $n \leq 8$, the $\pi$-bond model is completely inaccurate.

\begin{acknowledgements}
This work was supported by the Microelectronics Advanced Research
Corporation Focus Center on Nano Materials (FENA), SRC/SRCEA, the NSF (ECS-0524501), and
DARPA/DMEA-CNID (H94003-04-2-0404).
\end{acknowledgements}

\newpage

\newpage
\begin{center}
{\bf Figure Captions}
\end{center}

\noindent FIG. 1: One unit cell (10,0) zigzag CNT with
4 atomic layers. Self-energies take into account the
semi-infinite leads for transmission calculations.\\

\noindent FIG. 2: (a) $(n,0)$ CNT band gaps as a function of
$n$ and diameter
calculated from the
$\pi$-bond model and DFT with LDA and BLYP functionals.
(b) Difference between the band gap calculated from
the DFT models and the $\pi$-bond model.\\

\noindent FIG. 3. Electron, (a) - (c), and hole, (d) - (f), effective mass comparisons.
Top: Normalized effective mass $(m^*/m_0)$,
calculated from the $\pi$-bond and DFT models.
Middle: Difference between the effective mass calculated from
the DFT models and the $\pi$-bond model $(m_{\rm DFT} - m_{\pi -{\rm bond}}) / m_0$.
Bottom: Percent difference, $100*(m_{\rm DFT} - m_{\pi -{\rm bond}}) / m_{\pi -{\rm bond}}$.\\

\noindent FIG. 4. Transmission calculated from
$\pi$-bond and DFT models for $(n,0)$ CNTs with
$n$ values of (a) 10, (b) 20, (c) 31, and
(d) 35.\\

\newpage

\begin{center}
\begin{figurehere}
\includegraphics[width=3.5in]{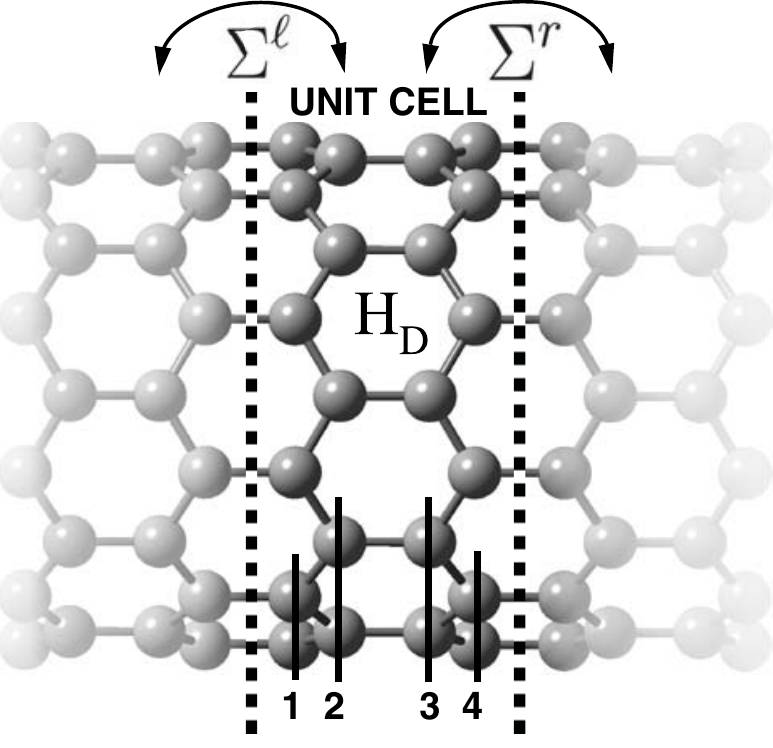}
\caption{} \label{fig:CNT_SCHEMATIC}
\end{figurehere}

\newpage
\begin{figurehere}
\includegraphics[width=3.5in]{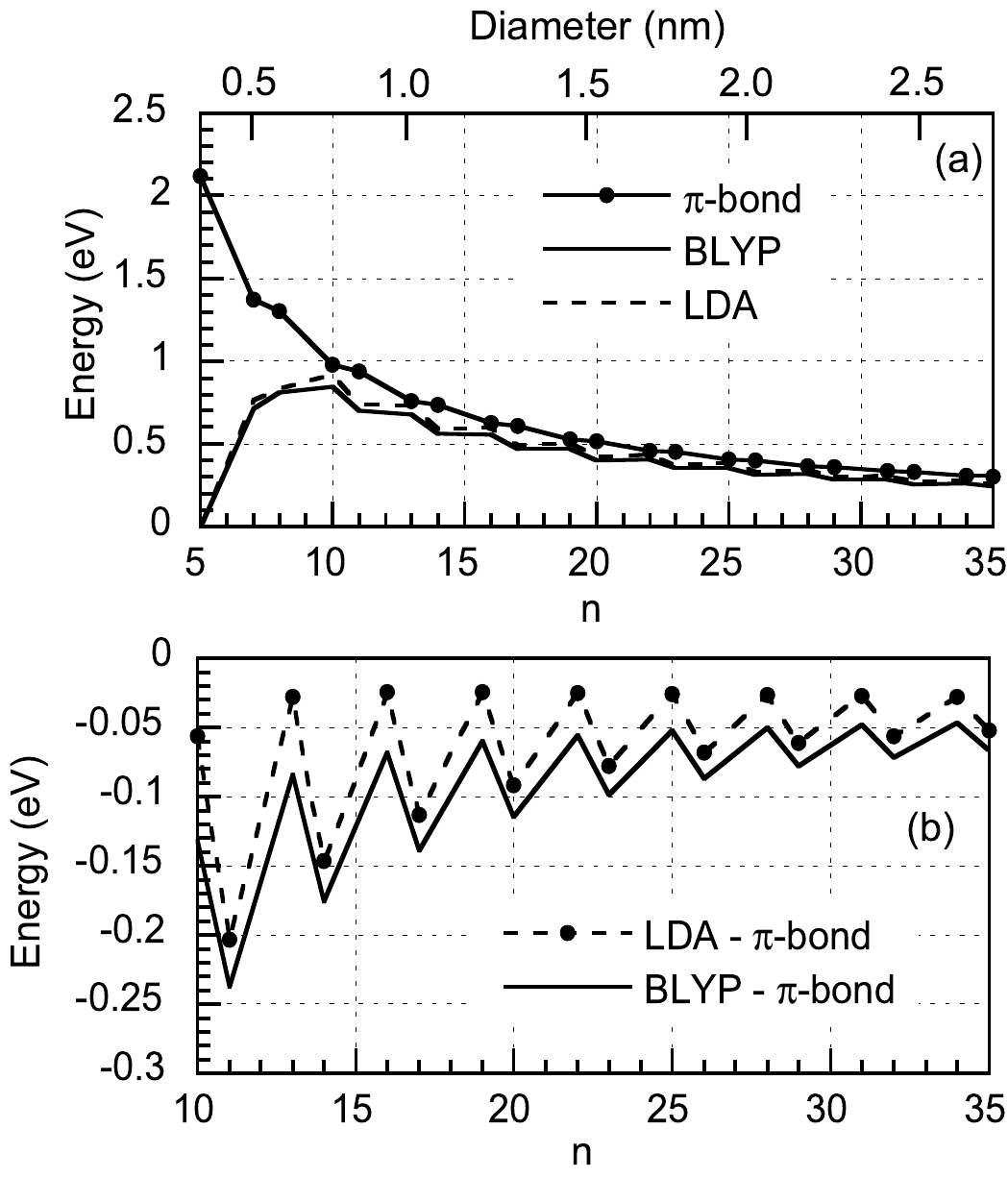}
\caption{} \label{fig:bandgap}
\end{figurehere}

\newpage
\begin{figurehere}
\includegraphics[width=6.0in]{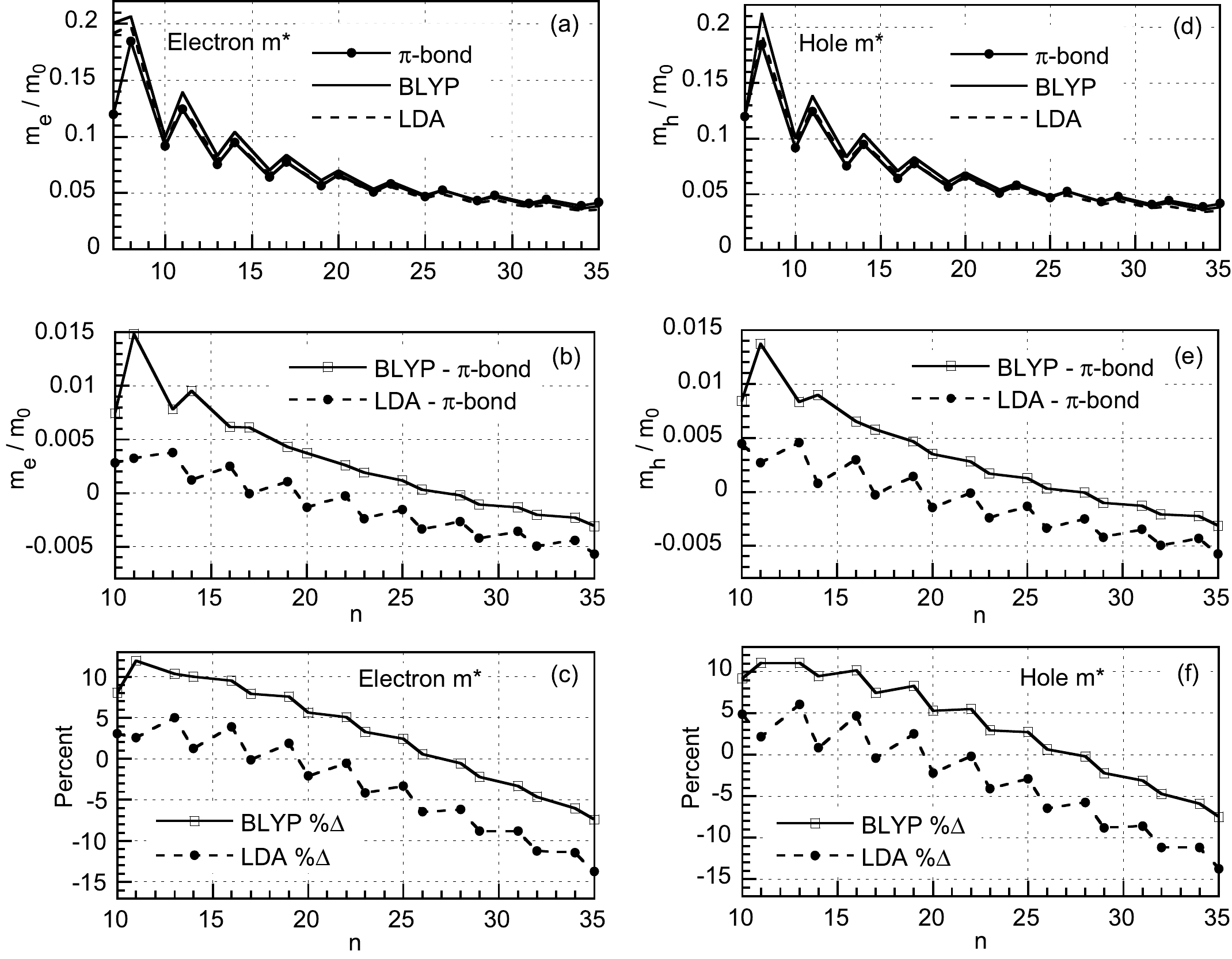}
\caption{} \label{fig:effM_comparison}
\end{figurehere}

\newpage
\begin{figurehere}
\includegraphics[width=6in]{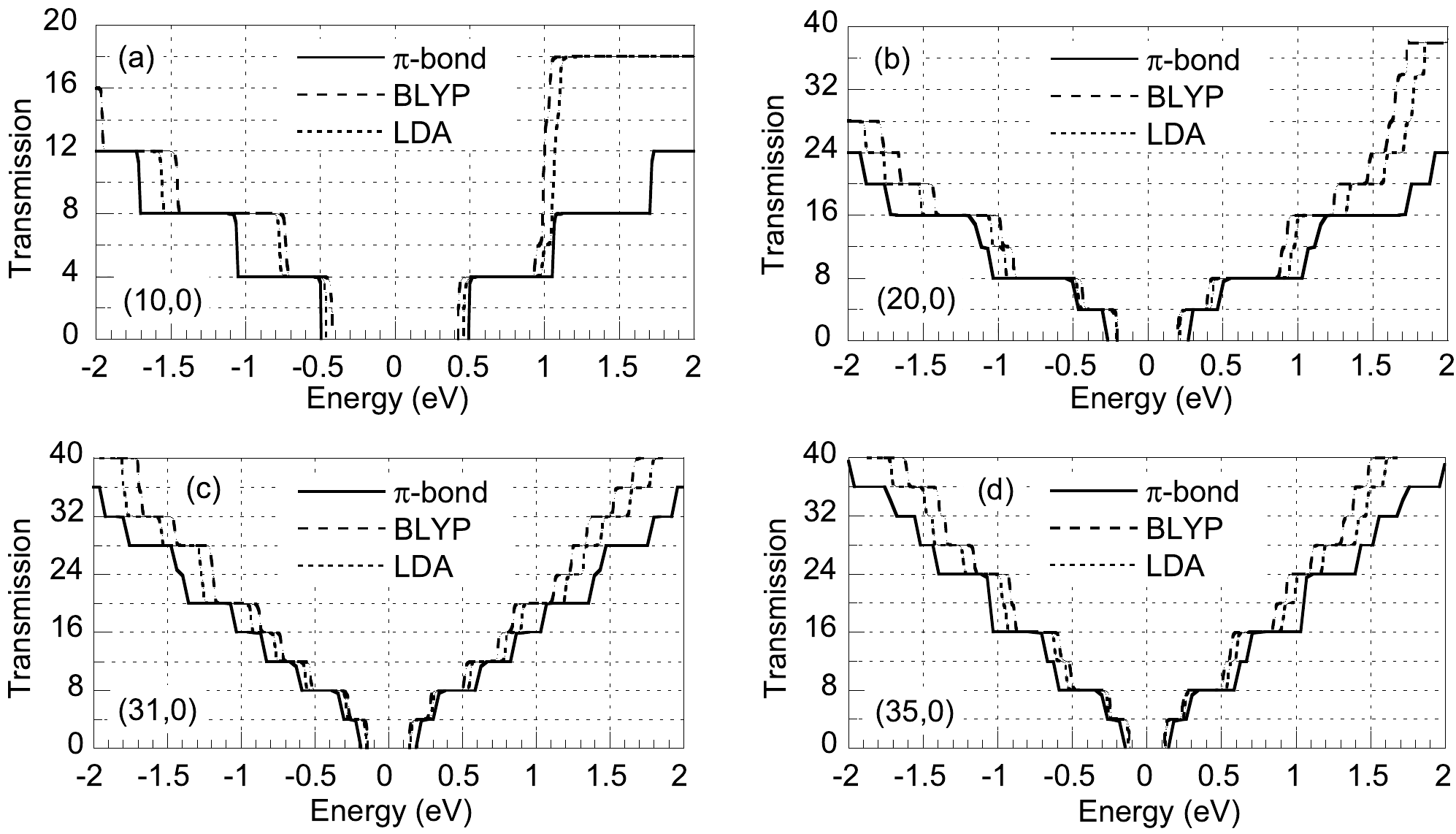}
\caption{} \label{fig:TE_comparison}
\end{figurehere}

\end{center}

\end{document}